\documentclass[letterpaper, 10 pt, conference]{ieeeconf}  
\makeatletter
\def\endthebibliography{%
  	\def\@noitemerr{\@latex@warning{Empty `thebibliography' environment}}%
  	\endlist
}
\makeatother

\IEEEoverridecommandlockouts                              
\usepackage{graphics} 
\usepackage{epsfig} 
\usepackage{mathptmx} 
\usepackage{times} 
\usepackage{amsmath} 
\usepackage{amssymb}  

\usepackage{amsthm}
\usepackage{xcolor}
\usepackage{tikz}
\usepackage{color}
\definecolor{mygreen}{rgb}{0.07, 0.35, 0.07}
\usepackage{float}

\newcommand\copyrighttext{%
	\footnotesize Copyright \textcopyright 2019 IEEE.  Personal use of this material is permitted.  Permission from IEEE must be obtained for all other uses, in any current or future media, including reprinting/republishing this material for advertising or promotional purposes, creating new collective works, for resale or redistribution to servers or lists, or reuse of any copyrighted component of this work in other works.}
\newcommand\copyrightnotice{%
	\begin{tikzpicture}[remember picture,overlay]
	\node[anchor=south,yshift=10pt] at (current page.south) {\fbox{\parbox{\dimexpr\textwidth-\fboxsep-\fboxrule\relax}{\copyrighttext}}};
	\end{tikzpicture}%
}

\title{\LARGE \bf
On the robust existence of piecewise quadratic Lyapunov functions for hybrid models of gene regulatory networks
}

\author{Mirko~Pasquini,~\IEEEmembership{Student Member,~IEEE,}
	David~Angeli,~\IEEEmembership{Fellow,~IEEE}
	\thanks{This work was supported by the Engineering and Physical Sciences Research Council (EPSRC) and the Department of Electrical and Electronic Engineering at Imperial College of London}
	\thanks{M. Pasquini is with the Department of Electrical and Electronic Engineering, Imperial College of London, UK. Email: {\tt\small mirko.pasquini16@imperial.ac.uk}}
	\thanks{D. Angeli is with the Department of Electrical and Electronic Engineering at the Imperial College of London, UK, and the Department of Information Engineering at the University of Florence, Italy. Email: {\tt\small d.angeli@imperial.ac.uk}}
}

\theoremstyle{definition}

\newtheorem{remark}{Remark}
\newtheorem{definition}{Definition}
\newtheorem{assumption}{Assumption}
\newtheorem{problem}{Problem}

\begin{document}

\maketitle
\copyrightnotice

\begin{abstract}
\textcolor{black}	
{	In this work we addressed the problem of stability analysis for an uncertain piecewise affine model of a genetic regulatory network. In particular we considered polytopic parameter uncertainties on the proteins production rate functions, giving conditions for the existence of a piecewise quadratic Lyapunov function for any possible realization of the system. In the spirit of other works in literature, the resulting conditions will be given on the vertices of the parameter polytope, while still taking into consideration the piecewise nature of the Lyapunov function and the presence, in general, of sliding modes solutions. An example is shown to prove the validity and applicability of the theoretical results.}
\end{abstract}

\color{black} 
\section{INTRODUCTION}
Due to surrounding environment and measurement noises, biological models often contain parameters that are uncertain and subject to variability \cite{Blanchini2014,Murray2010}, and for this reason it is appropriate to investigate the stability properties of such systems in a robust sense.\\  One of the most important concept in systems biology is that of Genetic Regulatory Networks (GRNs), as these networks describe the regulatory interactions between genes and proteins in a cellular environment \cite{Alon2007}.\\ In \cite{Casey2006} and \cite{DEJONG2004} the authors, using a piecewise affine (PWA) model of the dynamics of a GRN, employed a state transition graph (STG) -- which is unchanged in a large set of parameters -- to qualitatively describe the system trajectories. They found conditions that allow to determine stability properties of equilibria, which are related only to the STG.\\
In \cite{Pasquini2018}, we considered this PWA model and, using some interesting mathematical tools presented in \cite{Iervolino2017,Johansson1998}, we built an LMI framework that, solving a feasibility problem, gave the description of a piecewise quadratic (PWQ) Lyapunov function for the system. Nevertheless the solution found was strictly dependent on the system parameters, as these determine the values of the matrices defining the LMIs.\\ Many authors found ways to extend the Lyapunov approach, in order to deal with robustness stability when the system is subject to polytopic uncertainties. 
In \cite{Ramos2002} the authors considered a linear system, in which the system matrix belongs to a polytope of matrices. They gave a set of LMI conditions that, if verified, return the description of a Lyapunov function which is linearly dependent from the uncertain parameters, proving the Hurwitz stability of the entire set of system matrices in the aforementioned polytope. In \cite{Chesi2005} and \cite{Oliveira2006} the authors, considering the same kind of polytopic uncertainties, extended the approach considering Lyapunov functions that are dependent in an homogeneous polynomial way on the parameters. Despite considering linear systems with uncertain system matrices, these works did not consider switched systems and the problems associated with those, such as: the presence of sliding modes solutions and the continuity constraint to be asked in certain regions for a (generally) discontinuous piecewise quadratic Lyapunov function. In the following we will consider polytopic uncertainties on the production rate functions, in a PWA model of an N proteins system, and we present two approaches, in the spirit of \cite{Ramos2002,Chesi2005,Oliveira2006}, to prove the robust existence of a PWQ-LF under the aforementioned uncertainties, extending the LMI framework we described in \cite{Pasquini2018}. \\
The work is organized as follows: in Section \ref{sec:math_back} some mathematical background and notation are presented, while in Section \ref{sec:pwa_model} the piecewise affine model of the dynamics of a genetic regulatory network is briefly recalled. In Section \ref{sec:problem_formulation} the problem is formulated in a mathematically precise way and in Sections \ref{sec:common_approach} and \ref{sec:additional_constraint_approach} two solutions to the problem are illustrated. Finally Section \ref{sec:example} shows the application of the results to a numerical example while Section \ref{sec:conclusions} concludes the paper highlighting some possible further developments.

\section{NOTATION AND MATHEMATICAL BACKGROUND}
\label{sec:math_back}
Let $v = \begin{bmatrix} v_1 &\dots &v_n\end{bmatrix}^T$ be a vector in  $\mathbb{R}^n$. The notation $v \ge 0$ means that $ v_i \ge 0,\ \ \forall i \in \{1, \dots ,n\} $. With $\mathbb{R}^n_+ \subseteq \mathbb{R}^n$ we denote the set:
\begin{equation}
\mathbb{R}^n_+ := \{v \in \mathbb{R}^n \ | \ v \ge 0 \} 
\end{equation}
Let $\Omega := \{\omega_1, \dots, \omega_n\} \subset \mathbb{R}^n$ be a finite set of vectors in $\mathbb{R}^n$. The conic hull of $\Omega$ is defined as:
\textcolor{black}
{\begin{equation}
 cone\{\Omega\} := \{v \in \mathbb{R}^n \ | \exists \alpha \in \mathbb{R}^n_+ \text{ s.t. } v = \sum\limits_{i = 1}^n \alpha_i \omega_i\}
\end{equation}
}With $S_r$ we denote the standard simplex of dimension $r$, namely the set:
\begin{equation}
S_r := \{ v \in \mathbb{R}^r_+ \ | \ \sum\limits_{i=1}^r v_i = 1 \}
\end{equation}
The convex hull of $\Omega$ is defined as:
\begin{equation}
conv\{\Omega\} := \{v \in \mathbb{R}^n \ | \exists \alpha \in S_n \text{ s.t. } v = \sum\limits_{i = 1}^n \alpha_i \omega_i \}
\end{equation}
For any set $D$: $int(D)$ indicates the interior of $D$, $\partial D$ indicates its boundary and $cl(D) = int(D) \cup \partial D$ its closure. The notations $cl(D)$ and $\overline{D}$ can be used interchangeably.\\
Let $M \in \mathbb{R}^{n \times n}$ be a symmetric matrix of size $n$. $M$ is said to be positive definite (semidefinite), and indicated with $ M \succ 0$ ($M \succeq 0$), if:
\begin{equation}
x^T M x > 0 \ (x^T M x \ge 0) \qquad \forall x \in \mathbb{R}^n
\end{equation} 
$M$ is said to be negative definite (semidefinite), and indicated with $ M \prec 0$ ($M \preceq 0$), if $-M$ is positive definite (semidefinite).\\
A polyhedron $P$ in $\mathbb{R}^n$ is the set of solutions of a system of linear inequalities, formally:
\begin{equation}
\label{eq:H_representation_polyhedron}
P := \{x \in \mathbb{R}^n \ | \  A x \le b, \ A \in \mathbb{R}^{m \times n}, \ b \in \mathbb{R}^m \}
\end{equation}
The description \eqref{eq:H_representation_polyhedron} is called $H$-representation of $P$ and can always be converted in a $V$-representation \cite{Avis2002,MPT3}, formally:
\begin{equation}
P:= conv\{V\} + cone\{R\}
\end{equation}
where $V$ is the set of vertices of $P$ and $R$ the set of its rays. A procedure exist \cite{Iervolino2017} to associate an higher dimension cone, called homogenization cone, to any polyhedron $P$ and such cone will be used in the following to give conewise condition on quadratic functions.
\color{black}
\section{PIECEWISE AFFINE MODEL OF A GRN}
\label{sec:pwa_model}
\textit{\textcolor{black}{In this section we briefly recall the piecewise affine model considered for a genetic regulatory network. Most of this section is taken from \cite{Casey2006,Pasquini2018}, to which the reader is referred for a deeper discussion.}}\\
Consider a system of $n$ proteins and, for the single protein $X_i$, consider its concentration $x_i$ to evolve according to the dynamics:
\begin{equation}
\dot{x}_i = f_i(x) - c_i x_i, \qquad x_i \in \mathbb{R}_+
\end{equation}
in which \color{black} $x=\begin{bmatrix} x_1 &x_2 &\dots &x_n \end{bmatrix}^T$ \color{black} is the vector of proteins concentrations, $f_i(x) \ge 0$ describes the production rate of $X_i$ and $c_i > 0$ is its  degradation rate, that takes in consideration both diluition inside the cell and natural degradation of $X_i$. We assume $f_i(x)$ to be of the form:
\begin{equation}
\label{eq:production_rate}
f_i(x) := \sum_{j=1}^{l_i} b_{i,j} \cdot s_{i,j}(x) 
\end{equation}
for which $s_{i,j}(x)$ is a sum of products of step functions:
\begin{equation}
\label{eq:step_function_plus}
s^+(x_i,\theta_{i,k}) = \begin{cases} \begin{array}{lll} 1, &\text{ if } &x_i > \theta_{i,k}\\ 0, &\text{ if } &x_i < \theta_{i,k} \end{array} \end{cases}
\end{equation}
\begin{equation}
\label{eq:step_function_minus}
s^-(x_i,\theta_{i,k}) = 1 - s^+(x_i,\theta_{i,k})
\end{equation}
When considering all the $n$ proteins, we have the following model:
\begin{equation}
\label{eq:standard_vec_model}
\dot{x} = f(x) - C x, \qquad x \in \mathbb{R}^n_+
\end{equation}
in which $C \in \mathbb{R}^{n \times n}$ is a diagonal matrix with strictly positive diagonal entries and $f(x) : \mathbb{R}^n_+ \to \mathbb{R}^n_+$ is defined as:
\begin{equation}
\label{eq:production_rate_function_vector}
f(x) = \begin{bmatrix}
f_1(x) \\
\vdots \\
f_n(x)
\end{bmatrix}
\end{equation}
The structure of \eqref{eq:production_rate}  and the step functions \eqref{eq:step_function_plus} and \eqref{eq:step_function_minus}, naturally \color{black} partitions \color{black} the state space in boxes, defined by the thresholds $\theta$. Such boxes \color{black} (possibly with empty interior)  \color{black} will be called domains. If we consider a domain $D$, in which none of the $x_i$s assume a threshold value, such domain will be called regulatory, while if some $x_i$s assume a threshold value, the domain $D$ will be called switching. The set of regulatory domains is indicated with $\mathcal{D}_R$, while the set of switching domains is indicated with $\mathcal{D}_S$. \textcolor{black}{ For a domain $D_s \in \mathcal{D}_S$, we denote with $I_{D_s}$ the set of indexes of the $x_i$s that assume a threshold value in $D_s$.} \\ Inside a particular regulatory domain $D$ the function $f(x)$ is a constant vector and will be indicated with $f_D$, giving rise to the regulatory dynamics:
\begin{equation}
\label{eq:regulatory_dynamics}
\dot{x} = f_D - C x, \qquad x \in \mathbb{R}^n_+
\end{equation}
if $x \in D$.
However in a switching domain the function $f(x)$ is not properly defined in at least one of its components, so the system \eqref{eq:standard_vec_model} needs to be extended to a differential inclusion \cite{Filip88}, namely:
\begin{equation}
\label{eq:differential_inclusion_model}
\dot{x} \in H(x)
\end{equation}
with:
\begin{equation}
\label{eq:differential_inclusion_set_H}
H(x) = \begin{cases} \begin{array}{lll}
\{f_D - C x\}, &\text{if }  D \in \mathcal{D}_R\\
conv\{f_{D'} - C x \ | \ D' \in R(D)\}, &\text{if }  D \in \mathcal{D}_S
\end{array}
\end{cases}
\end{equation}
where $R(D)$ indicates the set of regulatory domains adjacent to the switching domain $D$.
In \cite{DEJONG2004,DeJong2003,Casey2006} the possibility to associate a State Transition Graph (STG) to the system is described. An STG is a graph in which: the nodes are associated to the domains of \eqref{eq:standard_vec_model}, while every edge is associated to the existence of a solution in the sense of Filippov \cite{Filip88}, which connects two domains, without crossing any other domain in between.\\
\textcolor{black}{In \cite{Pasquini2018} we developed an LMI framework that consents to find a Piecewise Quadratic (PWQ) Lyapunov function $V(x)$ for system \eqref{eq:standard_vec_model}. Such method is based on enforcing a set of LMIs and Matrix Equalities (MEs) constraints regarding: the continuity (only where required, in general the function will be discontinuous) and the monotonicity of $V$.\\
However, the feasibility of the problem is strictly related to the system parameters that, as stated at the beginning of this paper, are mostly uncertain and possibly varying with time. The next sections will deal with assuring the feasibility of the LMIs framework when the system is subject to polytopic uncertainties of the production rate function.} 

\section{PROBLEM FORMULATION}
\label{sec:problem_formulation}
Let $C \in \mathbb{R}^{n \times n}$ be a diagonal matrix with strictly positive diagonal entries.
Let $f^1(x)$, $\dots$, $f^L(x)$ be $L$ different production rate functions of the type \eqref{eq:production_rate_function_vector}. 
Let $\Sigma^k$ be the system:
\begin{equation}
\label{eq:external_system_equation}
\Sigma^k : \dot{x} = f^k(x) - C x, \qquad x \in \mathbb{R}^n_+
\end{equation}
\textcolor{black}{We call $\Sigma^k$ an extremal system. Let $\mathcal{C}_1^L$ be the set of systems:
\begin{equation}
\label{eq:C_1_L_set}
\mathcal{C}_1^L := \{ \sigma^\lambda : \dot{x} = f^\lambda - C x, \ x \in \mathbb{R}^n_+\}
\end{equation}
for which $f^\lambda$ has the form:
\begin{equation}
f^\lambda := \sum\limits_{k=1}^L \lambda_k f^k(x), \qquad \lambda \in S_L
\end{equation}
in which $S_L$ is the standard simplex of dimension $L$.}
Given the domain dependent nature of $f^k(x)$, any system $\sigma^\lambda \in \mathcal{C}_1^L$ has the dynamics:
\textcolor{black}
{\begin{equation}
\label{eq:parameters_variation_system_domain}
\dot{x} = \sum\limits_{k=1}^L \lambda_k f^k_{D} - C x	
\end{equation}
}inside the regulatory domain $D$.
We make the following assumption on $\mathcal{C}_1^L$.
\textcolor{black}
{\begin{assumption}
	\label{ass:same_stg}
	All the systems in $\mathcal{C}_1^L$ have the same state transition graph, the same thresholds and the same domains.$\hfill\blacksquare$
\end{assumption}
}
%
%
Our goal is to solve the following:
\begin{problem}
	\label{prob:Problem_paper}
	Find conditions under which any system in $\mathcal{C}_1^L$ admits a Piecewise Quadratic Lyapunov function. $\hfill\blacksquare$
\end{problem}
First we need to discuss how to describe sliding modes for a generic $\sigma^\lambda \in \mathcal{C}_1^L$, then we will explain two approaches to solve Problem \ref{prob:Problem_paper}, giving also a description of the resulting Lyapunov function.
\subsection{Sliding modes for systems in $\mathcal{C}_1^L$}
Let $\sigma^\lambda$ be a system in $\mathcal{C}_1^L$. We know that if $x \in D_S$, with $D_S$ being a switching domain, then $\sigma^\lambda$ should be extended to a differential inclusion \cite{Filip88}, and $\dot{x} \in H^\lambda(x)$, where $H^\lambda(x)$ is the set of directions $f$ satisfying the following equation:
\begin{equation}
\label{eq:directions_in_switching_lambda}
f = \sum\limits_{D' \in R(D_s)} \alpha_{D'} \left(\sum\limits_{k=1}^L \lambda_k f_{D'}^k\right) - C x
\end{equation}
with $\alpha \in S_{|R(D_s)|}$ and $\lambda \in S_L$. Equation \eqref{eq:directions_in_switching_lambda} can be written as:
\begin{equation}
f = \sum\limits_{D' \in R(D_s)} \sum\limits_{k=1}^L \alpha_{D'} \lambda_k f_{D'}^k - C x 
\end{equation}
and, given that both $\alpha$ and $\lambda$ belong to standard simplices, we have:
\begin{equation}
\begin{split}
\sum\limits_{D' \in R(D_s)} \sum\limits_{k=1}^L \alpha_{D'} \lambda_k = \sum\limits_{D' \in R(D_s)}  \alpha_{D'} \sum\limits_{k=1}^L  \lambda_k = \sum\limits_{D' \in R(D_s)} \alpha_{D'} = 1
\end{split}
\end{equation}
so that \eqref{eq:directions_in_switching_lambda} can be expressed as:
\begin{equation}
\label{eq:directions_in_switching_lambda_gamma}
f = \sum\limits_{D' \in R(D_s)} \sum\limits_{k=1}^L \gamma_{D'}^k f_{D'}^k - C x 
\end{equation}
with \textcolor{black} {$\gamma \in S_{L \cdot q}$} and $q := |R(D_s)|$.
 Following an approach similar to \cite{Pasquini2018}, the sliding mode directions of $\sigma^\lambda$ are certainly included in the ones described by \eqref{eq:directions_in_switching_lambda_gamma}, in which $\gamma$ belongs to the following polyhedron:
\begin{equation}
\label{eq:H_representation_of_the_polyhedron_P_gamma}
P_{\gamma,D_s} := \left\{\gamma \in \mathbb{R}^{L \cdot q} : 
\begin{cases}
\begin{array}{lcr}
\begin{bmatrix}\overline{F^1} &\overline{F^2} &\dots &\overline{F^L} \end{bmatrix} \gamma &=&  \begin{bmatrix}
\bar{c}\end{bmatrix} \\ 
1^T \gamma &=& 1\\
\gamma &\ge& 0 \\
\end{array} \!\!\!\!\!\!\!\!
\end{cases}\right\}
\end{equation}
\textcolor{black}{where $\bar{c}$ is a vector, the i-th component of which is:
\begin{equation}
[\bar{c}]_i = c_i \theta_{i,k}, \qquad  i \in I_{D_s}
\end{equation} where $c_i$ is the degradation rate of the protein $X_i$ and $\theta_{i,k}$ is the threshold value assumed by $x_i$ in the switching domain $D_s$, while $\overline{F^k}$ is the matrix:
\begin{equation}
\overline{F^k} = \begin{bmatrix}
(f^k_{D_{i_1}})^{I_{D_s}} &\dots& (f^k_{D_{i_q}})^{I_s}
\end{bmatrix}
\end{equation}
in which $(f^k_D)^{I_{D_s}}$ is the vector obtained by selecting only the components from $(f^k_D)$, indexed by the set of switching variables $I_{D_s}$ and $\{D_{i_1}, \dots, D_{i_q}\}$ is the set of regulatory domains adjacent to $D_S$.}
Because any $H$-representation of a polyhedron $P$ can be converted to its $V$-representation \cite{Avis2002,Iervolino2017},\textcolor{black} {and given the boundedness of $P_{\gamma,D_s}$}, we get:
\begin{equation}
P_{\gamma,D_s} := conv\{w_1,w_2, \dots, w_{\nu_\gamma}\}
\end{equation}

\begin{remark}
	\label{rem:sliding_mode_convex_hull}
	A direct consequence of the above analysis is that:
	\begin{equation}
	H^\lambda(x) \subseteq conv\{H^1(x), H^2(x), \dots, H^L(x)\}
	\end{equation}
	meaning that any element in $H^\lambda(x)$ is a convex combination of elements in the sets $H^1(x), H^2(x), \dots, H^L(x)$,\textcolor{black} { where $H^k(x)$ is the set \eqref{eq:differential_inclusion_set_H} for the extremal system $\Sigma^k$}. Nevertheless this does not guarantee that any sliding mode direction of $\sigma^\lambda$ is a convex combination of the sliding mode directions of the extremal systems $\Sigma^1$, $\dots$, $\Sigma^L$ (i.e. not all the directions in $H^k(x)$ are sliding mode directions for $\Sigma^k$).
	$\hfill\blacksquare$
\end{remark}
\section{Common Lyapunov function approach}
\label{sec:common_approach}
Intuitively if a common piecewise quadratic Lyapunov function $V$ exists for all the extremal systems $\Sigma^k, \ k \in \{1,\dots,L\}$, then $V$ is a possible Lyapunov function for any $\sigma^\lambda \in \mathcal{C}_1^L$. Unluckily, just searching for a common Lyapunov function will, almost certainly, result in an unfeasible problem, due to the fact that different systems in $\mathcal{C}_1^L$ will have different equilibria.
The following definition is instrumental to bypass this problem:
\begin{definition}
	A regulatory domain $D$ is said to be a sink domain if $\phi(D) \in D$.$\hfill\blacksquare$
\end{definition}
\textcolor{black}{We recall from \cite{DEJONG2004,Casey2006} that the focal point $\phi(D)$ of the regulatory domain $D$ for \eqref{eq:regulatory_dynamics} is defined as:
\begin{equation}
\phi(D) := C^{-1} f_D
\end{equation} }
Sink domains are the ones that drastically reduce the possibility of finding a common Lyapunov function for the extremal systems. On the other hand the dynamics inside sink domains is well characterized \cite{Casey2006}, with the property that any trajectory entering a sink domain will not leave it. This means that we don't need to actually define a Lyapunov function here, as we already know how the system \textcolor{black}{behaves. 
It is a direct consequence of Assumption \ref{ass:same_stg} that the set of sink domains is the same for all the systems in $\mathcal{C}_1^L$.}
The approach is then to search a piecewise quadratic function $V$, being a common Lyapunov function for the extremal systems, restricted only to non-sink domains.\\
Suppose that such function $V$ exists. We need to prove that $V$ is in fact a Lyapunov function for any $\sigma^\lambda \in \mathcal{C}_1^L$. 
 Inside a regulatory domain $D$ we have:
\begin{equation}
\begin{split}
	\dot{V}_D(x) &= {\nabla V_D(x)} \cdot \dot{x} \\&= {\nabla V_D(x)} \cdot \{\sum\limits_{k=1}^L \lambda_k (f_D^k - C x)\} \\ &= \sum\limits_{k=1}^L \lambda_k \left[{\nabla V_D(x)} \cdot  (f_D^k - C x)\right] \le 0
\end{split}
\end{equation} 
With respect to switching domains, by Remark \ref{rem:sliding_mode_convex_hull}, it is not sufficient for $V$ to be non-increasing along any sliding mode direction of any of the extremal systems, as some sliding mode direction of $\sigma^\lambda$ could be not included in those.
Let $D_S$ be a switching domain, for which $P_{\gamma,D_S}$ is non empty, and let $L_{D_s,j}$ be the matrix:
\begin{equation}
\label{eq:matrix_Lj_of_switching_domain_non_increasing}
L_{D_s,j}  = \begin{bmatrix} -2 P_D C & P_D F w_j - C d_D \\ w_j^T F^T P_D - (d_D)^T C & 2 (d_D)^T F w_j \end{bmatrix}
\end{equation}
in which $F$ is:
\begin{equation}
\label{eq:F_matrix_definition}
F := \begin{bmatrix}
F^1 &\dots &F^L
\end{bmatrix} 
\end{equation}
and $F^k$ is the matrix:
\begin{equation}
F^k := \begin{bmatrix}
 f^k_{D_{i_1}}&\dots& f^k_{D_{i_q}}
\end{bmatrix}
\end{equation} 
for the extremal system $\Sigma^k$.
If, together with the continuity of $V$ in $D_S$, the following set of LMIs is satisfied \textcolor{black} {\cite{Iervolino2017,Pasquini2018}}:
\begin{equation}
\label{eq:LMI_constraints_for_switching_domain_non_increasing_common}
\Gamma_{D_s}^T L_{D_s,j} \Gamma_{D_s} + M_{D_s,j} \preceq 0, \qquad \forall j \in \{1, \dots, \nu_\gamma\}
\end{equation}
in which $\Gamma_{D_s}$ is the ray matrix of the homogenization cone of $D_s$ and $M_{D_s,j}$ is any entrywise non-negative and symmetric matrix, then $V$ is non-increasing along any direction corresponding to $\gamma \in P_{\gamma,D_s}$ and because $P_{\gamma,D_s}$ contains the $\gamma$s that give rise to all the sliding mode directions of $\sigma^\lambda$ in $D_s$, this implies that $V$ is non-increasing along any sliding mode trajectories of $\sigma^\lambda$ in $D_S$. \\
\textcolor{black}{Lastly continuity constraints} on $V$ are dictated by the STG of the system, which is unchanged by Assumption \ref{ass:same_stg}, and by the switching domains $D_S$ for which the polyhedron $P_{\gamma,D_S}$ is non empty. Trivially, because of Assumption \ref{ass:same_stg} and the above discussion on sliding modes, such $V$ will respect all the required continuity constraints. 
The so found $V$ will then be a common piecewise quadratic Lyapunov function between all the systems in $\mathcal{C}_1^L$.
\section{Additional LMIs approach}
\label{sec:additional_constraint_approach}
Despite the simplicity of the approach in Section \ref{sec:common_approach}, the constraint of $V$ being in common between the extremal systems, can be too restrictive. Hence a second approach consists in finding $L$ different piecewise quadratic Lyapunov functions $V^1$, $V^2$, $\dots$, $V^L$ for the extremal systems that, together with the usual constraints described in \cite{Pasquini2018}, will satisfy a set of additional constraints, guaranteeing the existence of a function $V^\lambda$ for any $\sigma^\lambda \in \mathcal{C}_1^L$.\\
This approach is a generalization of the first one in the sense that if a solution exists to the problem defined in Section \ref{sec:common_approach}, then it will also be a feasible solution of the problem that we are now defining.\\
The function $V^\lambda$ is imposed to be a conic combination of $V^1$, $V^2$, $\dots$, $V^L$, formally:
\begin{equation}
\label{eq:V_lambda_function}
V^\lambda = \eta_1 V^1 + \eta_2 V^2 + \dots  + \eta_L V^L , \qquad \eta \in \mathbb{R}^L_+ 
\end{equation}
We call the functions $V^1$, $V^2$, $\dots$, $V^L$ extremal Lyapunov functions relative to the extremal systems $\Sigma^1, \Sigma^2, \dots, \Sigma^L$.
For clarity of the following explanation, the extremal Lyapunov function relative to the extremal system $\Sigma^k$ will be described by:
\begin{equation}
\label{eq:lyapunov_function_PWQ}
\begin{split}
V^k(x) &= V_D^k(x) \qquad \text{if } x \in D, \ D \in \mathcal{D}_R \\
V_D^k(x) &= x^T P_D^k x + 2 (d_D^k)^T x + \omega_D^k =\\ &= \begin{bmatrix} x^T &1 \end{bmatrix} \begin{bmatrix} P_D^k &d_D^k\\ (d_D^k)^T & \omega_D^k \end{bmatrix} \begin{bmatrix} x \\1 \end{bmatrix} = \bar{x}^T \overline{P_D^k} \bar{x}
\end{split}
\end{equation}
 In the subsequent discussion we are going to express all the constraints on $V^\lambda$ in terms of the extremal Lyapunov functions.
\subsection{Non-increasing in switching domains}
\label{sec:additional_constraint_approach_sliding}
The first set of constraints is relative to sliding mode solutions. In particular we need to ensure that the Lyapunov function $V^\lambda$ of $\sigma^\lambda$, is non-increasing along any sliding mode direction. 
To guarantee this we use a technique similar to the one applied in Section \ref{sec:common_approach}, asking that, in the switching domain $D_S$, all the extremal Lyapunov functions $V^k$ are non-increasing along any direction:
\begin{equation}
f = F \gamma
\end{equation}  
in which F is the one described in \eqref{eq:F_matrix_definition} and $\gamma \in P_{\gamma,D_S}$, with $P_{\gamma,D_s}$ described by \eqref{eq:H_representation_of_the_polyhedron_P_gamma}.
If the following set of LMIs is satisfied:
\begin{equation}
\label{eq:LMI_constraints_for_switching_domain_non_increasing}
\begin{split}
\Gamma_{D_s}^T L^k_{D_s,j} \Gamma_{D_s} + M^k_{D_s,j} \preceq 0, \qquad &\forall j \in \{1, \dots, \nu_\gamma\}\\
&\forall k \in \{1, \dots, L\}
\end{split}
\end{equation}
with:
\begin{equation}
\label{eq:matrix_Lj_of_switching_domain_non_increasing}
L^k_{D_s,j}  = \begin{bmatrix} -2 P^k_D C & P^k_D F w_j - C d_D \\ w_j^T F^T P^k_D - (d^k_D)^T C & 2 (d^k_D)^T F w_j \end{bmatrix}
\end{equation}
then the desired condition is satisfied as well.

\textcolor{black}{\subsection{Continuity constraints}
As discussed in Section \ref{sec:common_approach}, being the STG unchanged, and because of the constraints of Section \ref{sec:additional_constraint_approach_sliding}, the required continuity constraints are the same for all the systems in $\mathcal{C}_1^L$. 
For such reason all the functions $V^k$ are continuous in the same domains, and being the function $V^\lambda$ a conic combination of the functions $V^k$, it will be continuous in the same domains as well,
proving that however $\eta_1$, $\eta_2$, $\dots$, $\eta_L$ are chosen in \eqref{eq:V_lambda_function}, \textcolor{black} {all the required continuity constraints hold for $V^\lambda$.}}
\subsection{Non-increasing in regulatory domains}
The last set of constraints is related to the monotonicity of $V^\lambda$ inside regulatory domains. Let $D$ be a regulatory domain for the systems in $\mathcal{C}_1^L$. The extremal Lyapunov functions satisfy \cite{Pasquini2018}:
\begin{equation}
\label{eq:non_increase_regulatory_on_extremal_systems}
\begin{split}
\Gamma_D^T \tilde{P}_D^1 \Gamma_D &+ M_D^1 \preceq 0\\
\Gamma_D^T \tilde{P}_D^2 \Gamma_D &+ M_D^2 \preceq 0\\
&\dots \\
\Gamma_D^T \tilde{P}_D^L \Gamma_D &+ M_D^L \preceq 0\\
\end{split}
\end{equation}
where $\tilde{P}_D^k$ is the matrix:
\begin{equation}
\label{eq:P_tilde_D_k_matrix}
\tilde{P}_D^k  = \begin{bmatrix} -2 P_D^k C & P_D^k f_D^k - C d_D^k \\ (f_D^k)^T P_D^k - (d_D^k)^T C & 2 (d_D^k)^T f_D^k \end{bmatrix}
\end{equation}
and $M_D^k$ is a symmetric and entrywise-non-negative matrix.
Our goal is to give a set of constraints which guarantees that, for any $\sigma^\lambda \in \mathcal{C}_1^L$, it exists $\eta \in \mathbb{R}^L_+$ such that, for $V^\lambda$ as expressed in \eqref{eq:V_lambda_function}, the following LMI is satisfied:
\begin{equation}
\label{eq:LMI_non_increasing_reg_on_P_lambda}
\Gamma_D^T \tilde{P}_D^\lambda \Gamma_D + M_D^\lambda \preceq 0\\
\end{equation}
for any regulatory domain $D$, \textcolor{black}{ in which:
\begin{equation}
\tilde{P}_D^\lambda = \begin{bmatrix} -2 P_D^\lambda C & P_D^\lambda f_D^\lambda - C d_D^\lambda \\ (f_D^\lambda)^T P_D^\lambda - (d_D^\lambda)^T C & 2 (d_D^\lambda)^T f_D^\lambda\end{bmatrix}
\end{equation}
and:
\begin{equation}
\begin{split}
f_D^\lambda &:= \sum\limits_{k=1}^L \lambda_k f^k \\
P_D^\lambda &:= \sum\limits_{k=1}^L \eta_k P_D^k \\
d_D^\lambda &:= \sum\limits_{k=1}^L \eta_k d_D^k 
\end{split}
\end{equation}}
In order to do this the matrix $\tilde{P}_D^\lambda$ should be expressed in terms of the matrices $\tilde{P}_D^k$s.\\ 
Let $\left[\tilde{P}_D^\lambda\right]_{ij}$ denotes the block in position $(i , j)$ of $\tilde{P}_D^\lambda$ so that:
\begin{equation}
\label{eq:elements_of_P_tilde_D}
\begin{split}
	\left[\tilde{P}_D^\lambda\right]_{11} &= -2 (\sum\limits_{k=1}^L \eta_k P_D^k) C \\
	\left[\tilde{P}_D^\lambda\right]_{12} &= (\sum\limits_{k=1}^L \eta_k P_D^k)(\sum\limits_{k=1}^L \lambda_k f^k) - C \sum\limits_{k=1}^L \eta_k d_D^k \\
	\left[\tilde{P}_D^\lambda\right]_{21} &=\left[\tilde{P}_D^\lambda\right]_{12}^T\\
	\left[\tilde{P}_D^\lambda\right]_{22} &= 2(\sum\limits_{k=1}^L \eta_k d_D^k)(\sum\limits_{k=1}^L \lambda_k f^k)
\end{split}
\end{equation}
\textcolor{black}{If, respectively, $\left[\tilde{P}_D^k\right]_{ij}$ denotes the block in position $(i , j)$ of $\tilde{P}_D^k$, it is immediately clear that:}
\begin{equation}
\label{eq:element_1_1_of_P_tilde_D}
\left[\tilde{P}_D^\lambda\right]_{11} = \sum\limits_{k=1}^L (-2 \eta_k P_D^k C) = \sum\limits_{k=1}^L \eta_k \left[\tilde{P}_D^k\right]_{11}
\end{equation}
For the other terms more computations are needed. In particular for $\left[\tilde{P}_D^\lambda\right]_{12}$ it holds:
\begin{equation}
\label{eq:element_1_2_of_P_tilde_D_first}
\left[\tilde{P}_D^\lambda\right]_{12} = \sum\limits_{k=1}^L \sum\limits_{j=1}^L \eta_k \lambda_j P_D^k f^j  - C \sum\limits_{k=1}^L \eta_k d_D^k
\end{equation}
Knowing that $\lambda \in S_L$ we can rewrite \eqref{eq:element_1_2_of_P_tilde_D_first} as:
\begin{equation}
\label{eq:element_1_2_of_P_tilde_D_second}
\begin{split}
	\left[\tilde{P}_D^\lambda\right]_{12} &= \sum\limits_{k=1}^L \sum\limits_{\substack{j = 1 \\ j\neq k}}^L \eta_k \lambda_j P_D^k f^j  + \\ & + \sum\limits_{k=1}^L \eta_k (1- \sum\limits_{\substack{j = 1 \\ j\neq k}}^L \lambda_j) P_D^k f^k - C \sum\limits_{k=1}^L \eta_k d_D^k 
\end{split}
\end{equation}
Rearranging the terms of \eqref{eq:element_1_2_of_P_tilde_D_second} we get:
\begin{equation}
\label{eq:element_1_2_of_P_tilde_D_third}
\begin{split}
\left[\tilde{P}_D^\lambda\right]_{12} = \sum\limits_{k=1}^L \eta_k \left[\tilde{P}_D^k\right]_{12} + \\ + \sum\limits_{k=1}^L \sum\limits_{\substack{j = 1 \\ j\neq k}}^L (\eta_k \lambda_j P_D^k f^j - \eta_k \lambda_j P^k f^k ) 
\end{split}
\end{equation}
Defining:
\begin{equation}
\label{eq:delta_f_ij}
	\delta f^{kj} := f^j - f^k
\end{equation}
we obtain:
\begin{equation}
\label{eq:element_1_2_of_P_tilde_D_fourth}
\left[\tilde{P}_D^\lambda\right]_{12} = \sum\limits_{k=1}^L \eta_k \left[\tilde{P}_D^k\right]_{12} +  \sum\limits_{k=1}^L \sum\limits_{\substack{j = 1 \\ j\neq k}}^L \eta_k \cdot \lambda_j \cdot P_D^k \cdot \delta f^{kj} 
\end{equation}
Considering the following properties of $\delta f^{kj}$:
\begin{equation}
\label{eq:delta_f_ij_properties}
\begin{split}
\delta f^{kj} &= - \delta f^{jk} \\
\delta f^{kk} &= 0
\end{split}
\end{equation}
after a few manipulations of \eqref{eq:element_1_2_of_P_tilde_D_fourth} we obtain:
\begin{equation}
\label{eq:element_1_2_of_P_tilde_D_fifth}
\begin{split}
\left[\tilde{P}_D^\lambda\right]_{12} = \sum\limits_{k=1}^L \eta_k \left[\tilde{P}_D^k\right]_{12}  +\\+ \sum\limits_{k=1}^{L-1} \sum\limits_{j = k+1}^L (\eta_k \lambda_j  P_D^k - \eta_j \lambda_k P_D^j ) \cdot \delta f^{kj} 
\end{split}
\end{equation}
Now assume that $\lambda \in int(S_L)$ (i.e. $\lambda_k \ne 0, \forall k \in \{1,\dots,L\}$). If we choose $\eta_k$ as:
\begin{equation}
\label{eq:eta_choice}
\eta_k = {1 \over \prod\limits_{\substack{j = 1 \\ j \ne k}}^L \lambda_j }
\end{equation}
substituting \eqref{eq:eta_choice} in \eqref{eq:element_1_2_of_P_tilde_D_fifth}, and defining:
\begin{equation}
\label{eq:delta_f_ij}
\delta P_D^{kj} := P_D^j - P_D^k
\end{equation}
 gives:
\begin{equation}
\label{eq:element_1_2_of_P_tilde_D_sixth}
\begin{split}
\left[\tilde{P}_D^\lambda\right]_{12} = \sum\limits_{k=1}^L \eta_k \left[\tilde{P}_D^k\right]_{12}  - \sum\limits_{k=1}^{L-1} \sum\limits_{j = k+1}^L \xi_{kj} \cdot \delta P_D^{kj} \cdot \delta f^{kj} 
\end{split}
\end{equation}
in which:
\begin{equation}
\label{sigma_kj}
\xi_{kj} := \eta_k \lambda_j = \eta_j \lambda_k
\end{equation}
Moreover defining $\delta d_D^{kj} = d_D^j - d_D^k$, with similar reasoning we get:
\begin{equation}
\label{eq:element_2_2_of_P_tilde_D}
\left[\tilde{P}_D^\lambda\right]_{22} = \sum\limits_{k=1}^L \eta_k \left[\tilde{P}_D^k \right]_{22} - 2 \sum\limits_{k=1}^{L-1} \sum\limits_{j = k+1}^L \xi_{kj} \cdot (\delta d_D^{kj})^T \delta f^{kj} 
\end{equation}
\color{black}
If we put together \eqref{eq:element_1_1_of_P_tilde_D}, \eqref{eq:element_1_2_of_P_tilde_D_sixth} and \eqref{eq:element_2_2_of_P_tilde_D}, we can rewrite $P_D^\lambda$ as:
\begin{equation}
\label{eq:P_D_lambda_sum_of_matrices}
P_D^\lambda = \sum\limits_{k=1}^L \eta_k \tilde{P}_D^k +  \sum\limits_{k=1}^{L-1} \sum\limits_{j = k+1}^L \xi_{kj} \delta \tilde{P}_D^{kj}
\end{equation}
in which:
\begin{equation}
\label{eq:tilde_delta_P_D_expression}
\color{black} \delta \tilde{P}_D^{kj} \color{black} = \begin{bmatrix}
 0 & -\delta P_D^{kj}  \delta f^{kj} \\ -(\delta f^{kj})^T \delta P_D^{kj} & -2  (\delta d_D^{kj})^T \delta f^{kj} 
\end{bmatrix}
\end{equation}
\color{black}
Using \eqref{eq:P_D_lambda_sum_of_matrices}, condition \eqref{eq:LMI_non_increasing_reg_on_P_lambda} can be written as:
\begin{equation}
\label{eq:LMI_for_lambda_V}
\Gamma_D^T  \left( \sum\limits_{k=1}^L \eta_k \tilde{P}_D^k +  \sum\limits_{k=1}^{L-1} \sum\limits_{j = k+1}^L \xi_{kj} \delta \tilde{P}_D^{kj} \right)\Gamma_D + M_D^\lambda \preceq 0
\end{equation}
and rearranging the terms in \eqref{eq:LMI_for_lambda_V} we obtain:
\begin{equation}
\label{eq:LMI_for_Lambda_V_final_form}
\sum\limits_{k=1}^L \eta_k \Gamma_D^T \tilde{P}_D^k \Gamma_D +  \sum\limits_{k=1}^{L-1} \sum\limits_{j = k+1}^L \xi_{kj} \Gamma_D^T \delta \tilde{P}_D^{kj} \Gamma_D  + M_D^\lambda \preceq 0
\end{equation}
Because constraints \eqref{eq:non_increase_regulatory_on_extremal_systems} hold and because any $\xi_{kj} > 0$, then satisfying the following set of LMIs:
\begin{equation}
\label{eq:constraint_to_robust_existence_regulatory}
\begin{split}
\Gamma_D^T \delta \tilde{P}_D^{kj} \Gamma_D + M^{kj} \preceq 0, \qquad &k \in \{1,\dots,L-1\} \\ &j \in \{i+1,\dots,L\}
\end{split}
\end{equation}
with $M^{kj}$ being an entrywise non-negative and symmetric matrix, is sufficient to guarantee that \eqref{eq:LMI_non_increasing_reg_on_P_lambda} is satisfied.
\begin{remark}
	Being:
	\begin{equation}
	\eta_k = {1 \over \prod\limits_{\substack{j = 1 \\ j \ne k}}^L \lambda_j }
	\end{equation}
	the Lyapunov function $V^\lambda$ for $\sigma^\lambda$ is described by:
	\begin{equation}
	V^\lambda := \sum\limits_{k=1}^L {1 \over \prod\limits_{\substack{j = 1 \\ j \ne k}}^L \lambda_j } V^k
	\end{equation}
	Considering $\lambda \in int(S_L)$, and considering that $\alpha V^\lambda$ is still a Lyapunov function for $\sigma^\lambda$, $\forall \alpha > 0$, we have that:
	\begin{equation}
	\hat{V}^\lambda := \sum\limits_{k=1}^L  \lambda_k V^k
	\end{equation}
	is a Lyapunov function for $\sigma^\lambda$. $\hfill\blacksquare$
\end{remark}
\begin{remark}
	The analysis done in the Section \ref{sec:additional_constraint_approach} has been done for $\lambda \in int(S_L)$. Anyway if $\lambda \in \partial S_L$ it is possible to prove that the problem reduces to a lower dimensional one -- meaning that if, for example, one of the $\lambda_k = 0$, the problem is equivalent to a problem with $L-1$ extremal systems instead of L -- and the relative constraints are contained in the ones already asked.
	$\hfill\blacksquare$
\end{remark}
\begin{remark}
	As in Section \ref{sec:common_approach}, if the constraints relative to sink domains are dropped, the possibility to find a feasible solution drastically improves. 	$\hfill\blacksquare$
\end{remark}
\section{EXAMPLE}
\label{sec:example}
\subsection{System with sliding mode}
\begin{figure}[H]
	\centering
	\includegraphics[width=8.5cm]{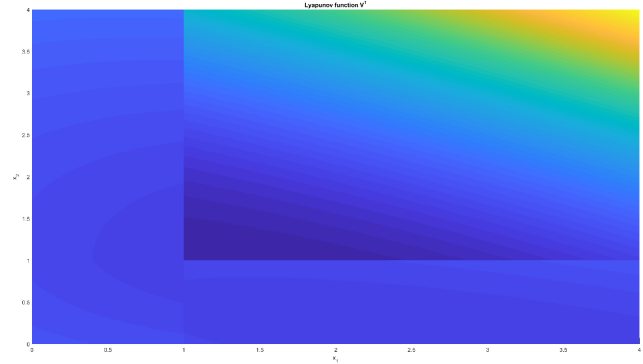}
	\caption{Extremal Lyapunov function $V^1$}
	\label{fig:extremal_V1_sliding}
\end{figure}

\begin{figure}[H]
	\centering
	\includegraphics[width=8.5cm]{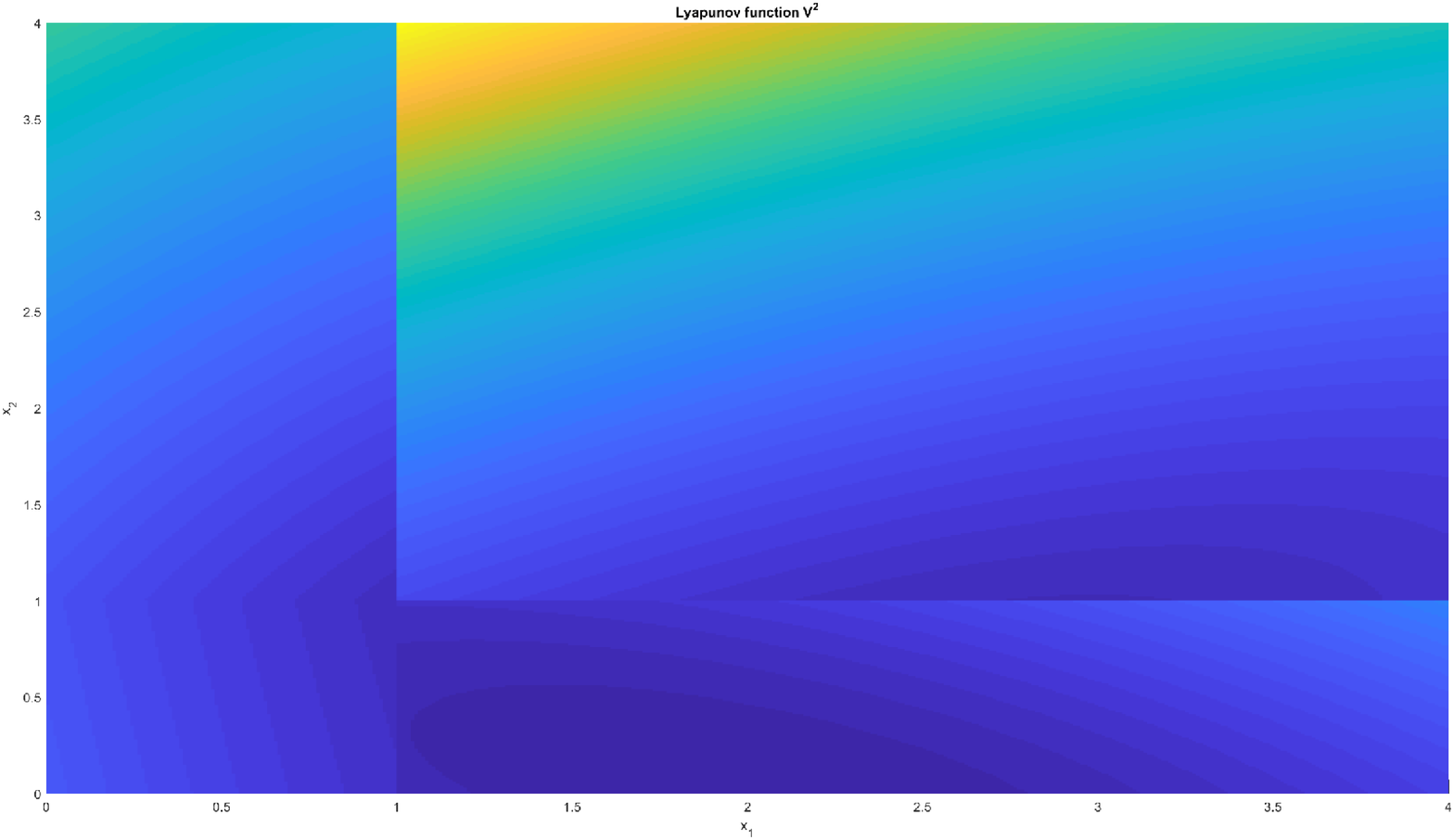}
	\caption{Extremal Lyapunov function $V^2$}
	\label{fig:extremal_V2_sliding}
\end{figure}

\begin{figure}[H]
	\centering
	\includegraphics[width=8.5cm]{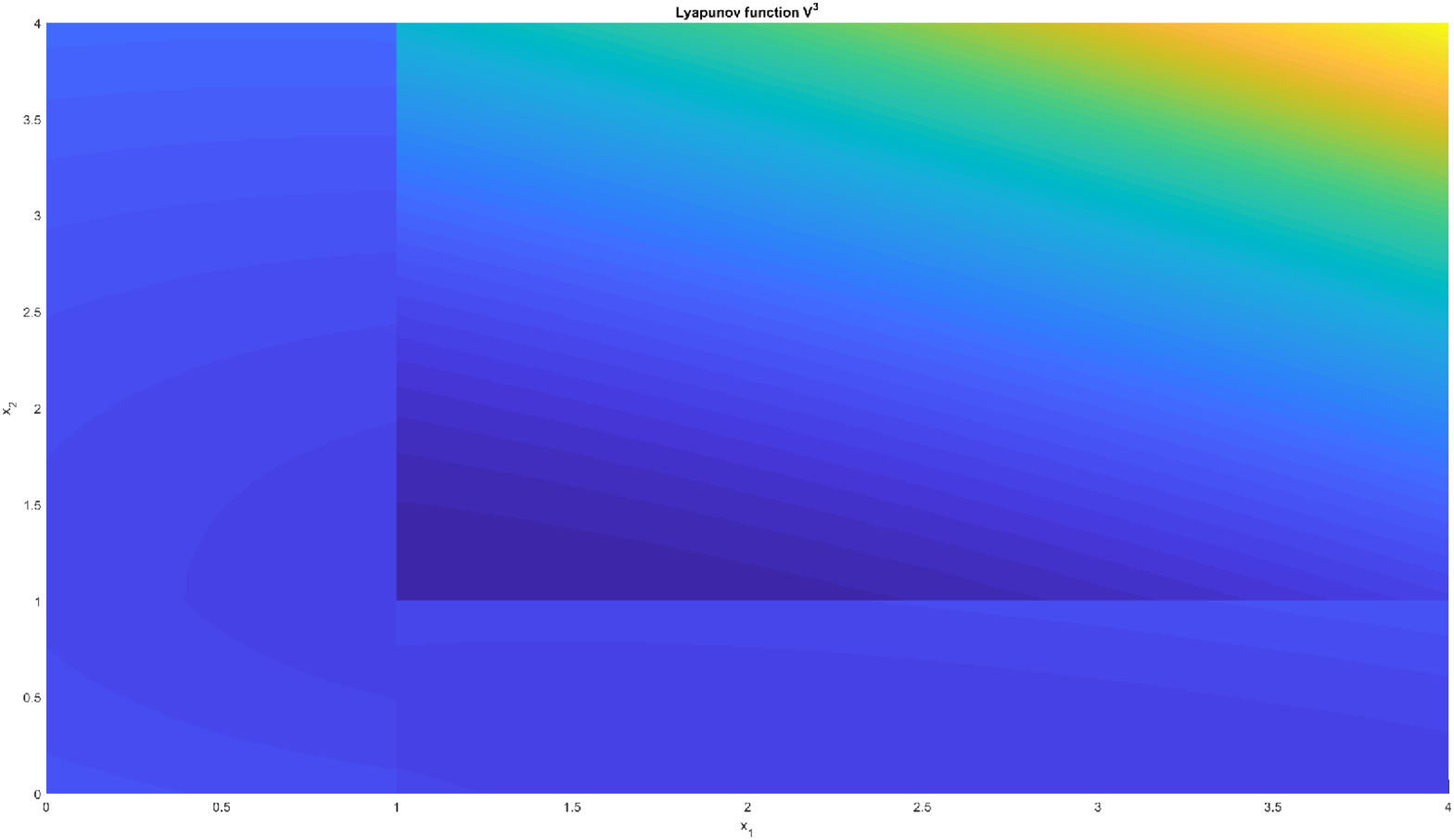}
	\caption{Extremal Lyapunov function $V^3$}
	\label{fig:extremal_V3_sliding}
\end{figure}

\begin{figure}[H]
	\centering
	\includegraphics[width=8.5cm]{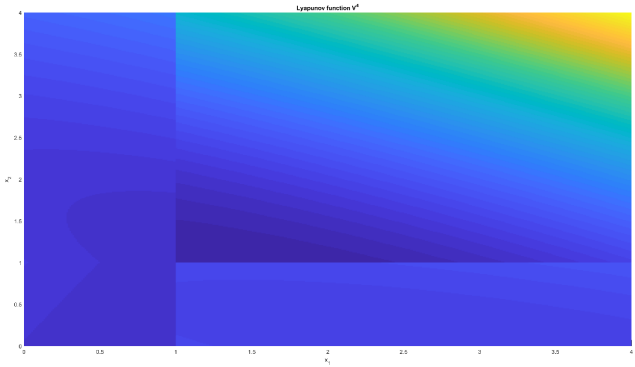}
	\caption{Extremal Lyapunov function $V^4$}
	\label{fig:extremal_V4_sliding}
\end{figure}

\begin{figure}[H]
	\centering
	\includegraphics[width=8.5cm]{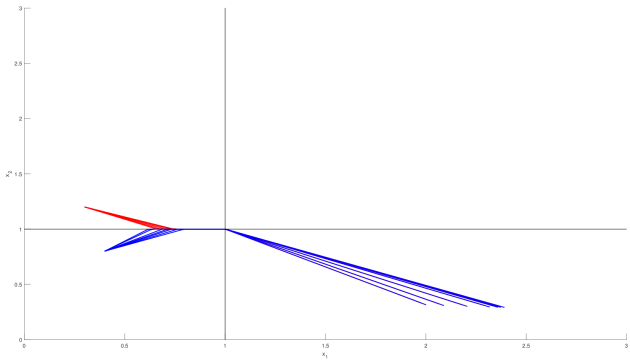}
	\caption{System trajectories for different $\lambda$ from two different initial points.}
	\label{fig:trajectories_sliding}
\end{figure}

\begin{figure}[H]
	\centering
	\includegraphics[width=8.5cm]{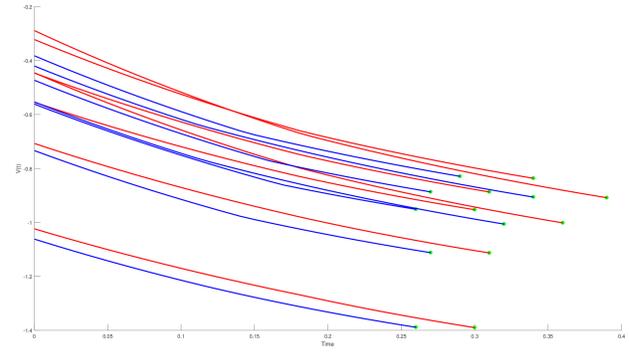}
	\caption{Evolution in time of $V^\lambda$ for different values of $\lambda$. The green dots represent the points where the trajectories  enter the domain $D_2 := (1,\infty) \times (0,1)$, which is a sink domain and for this reason is not included in the feasibility problem.}
	\label{fig:V_lambda_sliding}
\end{figure}

Consider the system:
\begin{equation}
\label{eq:sliding_system}
\begin{cases}   
\dot{x}_1 = 2 \cdot (\lambda_1 + \lambda_2) + 3 \cdot(\lambda_3 + \lambda_4) -  x_1\\
\dot{x}_2 = \left[2 \cdot (\lambda_1 + \lambda_3) +3 \cdot (\lambda_2 + \lambda_4)\right] s(x) -  x_2
\end{cases}
\end{equation}
where:
\begin{equation}
s(x) := s^-(x_1,1)s^-(x_2,1)
\end{equation}
We want to guarantee the existence of a Lyapunov function for $\lambda \in S_4$.  
We apply the approach of Section \ref{sec:additional_constraint_approach} so, given a system $\sigma^\lambda \in \mathcal{C}_1^L$, we define a set of additional constraints on the extremal Lyapunov functions, guaranteeing that their convex combination with weights $\lambda_k$ is a Lyapunov function for $\sigma^\lambda$. The algorithm is able to find a feasible solution in the extremal Lyapunov functions showed in Figures \ref{fig:extremal_V1_sliding}--\ref{fig:extremal_V4_sliding}.
For the system $\sigma^\lambda$, the function:
\begin{equation}
V^\lambda = \lambda_1 V^1 + \lambda_2 V^2 + \lambda_3 V^3 + \lambda_4 V^4
\end{equation}
is then a Lyapunov function.
To test the validity of the results we generate values of $\lambda \in S_4$ and evaluate the function $V^\lambda$ (Figure \ref{fig:V_lambda_sliding}) along the trajectories of system \eqref{eq:sliding_system} (Figure \ref{fig:trajectories_sliding}).
\textcolor{black}
{\section{CONCLUSIONS AND FUTURE WORK}
\label{sec:conclusions}
We considered a piecewise affine model of the dynamics of a GRN, subject to polytopic uncertainties of the production rate function. We addressed the problem of finding conditions that guarantee the existence of a piecewise quadratic Lyapunov function for any of the systems in the set $\mathcal{C}_1^L$, the dynamics of which is obtained through a convex combination of the dynamics of a finite number of, what we called, extremal systems. We proposed two solutions to this problem: the first consists in searching for a common Lyapunov function among the extremal systems, limited only to particular domains, while the second approach is a generalization of the first one and consists in searching for different Lyapunov functions for all the extremal systems that, when linearly combined, can generate Lyapunov functions for all the systems in $\mathcal{C}_1^L$. To conclude we proposed a numerical example, to show the validity of the described approach.\\
Future works should focus on the properties shared by the Lyapunov functions for the systems in $\mathcal{C}_1^L$, aiming to prove a result \textcolor{black} {of robust convergence for the system. }}

\bibliography{mirkos_bibliography}

\begin{thebibliography}{10}
\providecommand{\url}[1]{#1}
\csname url@samestyle\endcsname
\providecommand{\newblock}{\relax}
\providecommand{\bibinfo}[2]{#2}
\providecommand{\BIBentrySTDinterwordspacing}{\spaceskip=0pt\relax}
\providecommand{\BIBentryALTinterwordstretchfactor}{4}
\providecommand{\BIBentryALTinterwordspacing}{\spaceskip=\fontdimen2\font plus
\BIBentryALTinterwordstretchfactor\fontdimen3\font minus
  \fontdimen4\font\relax}
\providecommand{\BIBforeignlanguage}[2]{{%
\expandafter\ifx\csname l@#1\endcsname\relax
\typeout{** WARNING: IEEEtran.bst: No hyphenation pattern has been}%
\typeout{** loaded for the language `#1'. Using the pattern for}%
\typeout{** the default language instead.}%
\else
\language=\csname l@#1\endcsname
\fi
#2}}
\providecommand{\BIBdecl}{\relax}
\BIBdecl

\bibitem{Blanchini2014}
F.~Blanchini and G.~Giordano, ``{Piecewise-linear Lyapunov functions for
  structural stability of biochemical networks},'' \emph{Automatica}, vol.~50,
  no.~10, pp. 2482--2493, Oct 2014.

\bibitem{Murray2010}
\BIBentryALTinterwordspacing
R.~M. Murray and D.~Del~Vecchio, \emph{Biomolecular Feedback Systems}.\hskip
  1em plus 0.5em minus 0.4em\relax Princeton University Press, 2014. [Online].
  Available: \url{http://www.cds.caltech.edu/~murray/BFSwiki}
\BIBentrySTDinterwordspacing

\bibitem{Alon2007}
U.~Alon, \emph{An Introduction to Systems Biology: Design Principles of
  Biological Circuits}, ser. Mathematical and Computational Biology
  Series.\hskip 1em plus 0.5em minus 0.4em\relax Chapman \& Hall /CRC, 2007,
  vol.~10.

\bibitem{Casey2006}
R.~Casey, H.~{De Jong}, and J.~L. Gouz{\'{e}}, ``{Piecewise-linear models of
  genetic regulatory networks: Equilibria and their stability},'' \emph{Journal
  of Mathematical Biology}, vol.~52, no.~1, pp. 27--56, Jan 2006.

\bibitem{DEJONG2004}
H.~De~Jong, J.~Gouz{\'{e}}, C.~Hernandez, M.~Page, T.~Sari, and J.~Geiselmann,
  ``Qualitative simulation of genetic regulatory networks using
  piecewise-linear models,'' \emph{Bulletin of Mathematical Biology}, vol.~66,
  no.~2, pp. 301--340, Mar 2004.

\bibitem{Pasquini2018}
M.~Pasquini and D.~Angeli, ``On piecewise quadratic lyapunov functions for
  piecewise affine models of gene regulatory networks,'' in \emph{2018 IEEE
  57th Annual Conference on Decision and Control, CDC 2018}, Miami Beach, FL,
  USA, Dec 2018, pp. 1071--1076.

\bibitem{Iervolino2017}
R.~Iervolino, D.~Tangredi, and F.~Vasca, ``Lyapunov stability for piecewise
  affine systems via cone-copositivity,'' \emph{Automatica}, vol.~81, pp.
  22--29, Jul 2017.

\bibitem{Johansson1998}
M.~Johansson and A.~Rantzer, ``{Computation of Piecewise Quadratic Lyapunov
  Functions for Hybrid Systems},'' \emph{IEEE Transactions on Automatic
  Control}, vol.~43, no.~4, Apr 1998.

\bibitem{Ramos2002}
D.~C.~W. Ramos and P.~L.~D. Peres, ``{An LMI condition for the robust stability
  of uncertain continuous-time linear systems},'' \emph{IEEE Transactions on
  Automatic Control}, vol.~47, no.~4, pp. 675--678, Apr 2002.

\bibitem{Chesi2005}
G.~Chesi, A.~Garulli, A.~Tesi, and A.~Vicino, ``{Polynomially
  parameter-dependent Lyapunov functions for robust stability of polytopic
  systems: An LMI approach},'' \emph{IEEE Transactions on Automatic Control},
  vol.~50, no.~3, pp. 365--370, Mar 2005.

\bibitem{Oliveira2006}
R.~C. L.~F. Oliveira and P.~L.~D. Peres, ``{LMI conditions for robust stability
  analysis based on polynomially parameter-dependent Lyapunov functions},''
  \emph{Systems and Control Letters}, vol.~55, no.~1, pp. 52--61, Jan 2006.

\bibitem{Avis2002}
D.~Avis, K.~Fukuda, and S.~Picozzi, ``On canonical representations of convex
  polyhedra,'' in \emph{Proceedings of the First International Congress of
  Mathematical Software}, 2002, pp. 350--360.

\bibitem{MPT3}
M.~Herceg, M.~Kvasnica, C.~Jones, and M.~Morari, ``{Multi-Parametric Toolbox
  3.0},'' in \emph{Proc.~of the European Control Conference}, Z\"urich,
  Switzerland, July 2013, pp. 502--510.

\bibitem{Filip88}
A.~Filippov, \emph{Differential Equations with Discontinuous Righthand Sides},
  F.~Arscott, Ed.\hskip 1em plus 0.5em minus 0.4em\relax Kluwer Academic
  Publishers, 1988.

\bibitem{DeJong2003}
H.~De~Jong, J.~Geiselmann, C.~Hernandez, and M.~Page, ``{Genetic Network
  Analyzer: qualitative simulation of genetic regulatory networks},''
  \emph{Bioinformatics}, vol.~19, no.~3, pp. 336--344, Feb 2003.

\end{thebibliography}
\bibliographystyle{IEEEtran}
\end{document}